\documentclass[11pt]{article}
\usepackage{graphicx}
\begin{document}
{\Large Scaling in a simple model for surface growth in a random
medium}

\bigskip
Amnon Aharony$^1$ and Dietrich Stauffer$^{1,2}$

\bigskip
\noindent
$^1$ School of Physics and Astronomy, Raymond and Beverly Sackler Faculty of
Exact Sciences, Tel Aviv University, Ramat Aviv, Tel Aviv 69978, Israel

\centerline {e-mail: aharony@post.tau.ac.il}

\medskip
\noindent
$^2$ Institute for Theoretical Physics, Cologne
University, D-50923 K\"oln,  Euroland

\centerline {e-mail: stauffer@thp.uni-koeln.de}

\bigskip
{\small  Surface growth in random media is usually governed by
both the surface tension and the random local forces. Simulations
on lattices mimic the former by imposing a maximum gradient $m$ on
the surface heights, and the latter by site-dependent random
growth probabilities. Here we consider the limit $m \rightarrow
\infty$, where the surface grows at the site with minimal random
number, {\it independent} of its neighbors.  The resulting height
distribution obeys a simple scaling law, which is destroyed when
local surface tension is included. Our model is equivalent to
Yee's simplification of the Bak-Sneppen model for the extinction of
biological species, where the height represents the number of
times a biological species is exchanged. }
\bigskip


\bigskip
Invasion percolation \cite{WW} is a model for viscous fingering,
in which at each time step a segment of the interface moves into
the capillary channel which imposes the (overall) minimum
resistance. Numerically, this is modeled by a lattice of sites,
where each site is assigned a random number between zero and one,
and by moving the interface into the perimeter site with the
smallest random number. After a while, most of the perimeter
random numbers are relatively large (above the percolation
threshold), and growth occurs in ``bursts" which explore the
vicinity of where they started \cite{furuberg}.

Another family of growth models is based on the ``solid on solid"
concept, where one ignores overhangs and allows only steps which
increase the height $h$ of the interface (relative to the initial
line or plane). Such models, which also allow for surface tension,
include the Kardar-Parisi-Zhang (KPZ) \cite{KPZ} model, where the
random resistance depends on horizontal position $x$ and on time,
and the Parisi \cite{parisi} model, where this resistance is
local, depending on both $x$ and $h$. A discrete version of the
latter, by Kim and Kosterlitz \cite{KK}, only allows steps which
obey the constraint $|h(x)+1-h(x \pm 1)| \le m$, with $m=1$.
Sneppen (and Jensen) \cite{Sneppen} considered variants of this
model, in which neighbors continue to be updated until the
Kim-Kosterlitz constraint is obeyed everywhere. Bak and Sneppen
\cite{BS} then used a related model to describe the extinction of
biological species. In their model, they always updated also the
nearest neighbors of the growing site, irrespective of the height
gradients. However, unlike Sneppen and Jensen, they did not
continue these updates in an ``avalanche" beyond the nearest
neighbors.  The ``bursts" were thus more localized than in the
original Sneppen model.

Recently, Yee \cite{Yee} introduced a simplified version of the
Bak-Sneppen model, which can be identified as the $m \rightarrow
\infty$ limit. In this model, there is no surface tension
constraint on the growth, and thus each ``pillar" continues to
grow as long as the random number in front of it is smaller than
those facing all other pillars. Although this model preserves the
long range information on all the perimeter ``resistances", it
omits the interactions between neighbors, and is thus geometry and
dimension-independent. In the language of invasion percolation,
this model could describe a collection of independent capillary
channels, penetrated from one end by a viscous fluid from the {\it
same} reservoir, and with independent randomly varying resistances
along each channel. The removal of interactions allowed an
analytic solution of many aspects of this model \cite{SN}, in
agreement with simulations, which were phrased in the biological
language. In the present paper we re-interpret this model as one
of surface growth, and test if the resulting height distribution
of this growth obeys a standard scaling law. We start with the
definition of the model and its simulation and end with a partial
analytical treatment.

The Yee version of the Bak-Sneppen model takes an array of $L$
random numbers $r_i, \; i=1,2,\dots,L$ initially distributed
between zero and unity. At each time step $t \rightarrow t+1$, the
smallest of the $L$ random numbers is replaced by a new random
number. After some time \cite{SN}, nearly all random numbers are
very close to unity, and growth occurs in large ``bursts" where a
single ``pillar" grows, similar to invasion percolation
\cite{furuberg}. This version of the model \cite{Yee} is
independent of any geometry; the full Bak-Sneppen model replaces
also the lattice neighbors of the lowest random number, and thus depends
on the assumed lattice geometry; it gives a threshold
$x_c$ below unity such that after a long time $x_c < r_i < 1$ for
nearly all $i$. The random numbers $r_i$ can be interpreted as
fitness \cite{BS} in biological evolution, or as the quality of
court decisions in a judicial system with law-by-precedent
\cite{Yee}. We now use the physical interpretation \cite{Sneppen}.

With every element $i$ we associate a height variable $h_i$ which
initially is zero for all $i$, and then increases by unity every
time this element $i$ gets a new random value $r_i$ (which
corresponds to the new perimeter site in front of the moving
interface). We can imagine a deposition process in which bricks
drop down onto the site with the lowest random number $r_i$ and
then change this $r_i$ into another random number, or as the
viscous fluid moving along the $i$'th (one dimensional) capillary
channel, and reaching a new resistance. What is the probability
distribution function (proportional to the histogram) of the observed
heights? For this purpose we stop the growth whenever the highest
$h_i$ value reaches a predetermined value $L_z$ (as usual in
invasion percolation simulations and experiments \cite{furuberg}).

To avoid overcrowding of our figures we binned the observed
heights into powers of two; that means the $k$-th bin contained
heights between $2^{k-1}$ and $2^k-1$. It is plausible that for $1
\ll h_i \le L_z$ and large $L_z$ the results should depend mainly
on the ratio $h_i/L_z$. Figure 1 shows in its three parts the
binned histogram $N(h)$ in the scaled form $N/L_z$ versus
$h/L_z$, giving a good data collapse for large enough $L$; the
three parts correspond to $L_z/L$ = 0.1, 1 and 10 and for large
$L$ seem to give the same curve. The initial linear increase, for
$h \ll L$, in these log-log plots, together with the fact that the
bin size increases as the height $h$, would imply that without
binning this increase corresponds to a constant probability
distribution function $\propto N$. At large heights we see a cut-off since
$h > L_z$ is impossible. In fact, Fig. 2 shows that the unbinned
distribution follows roughly an exponential, $\propto \exp(-6h/L$).

The time $\tau$ after which the tallest pillar hits the top,
$h_i=L$, is shown in Fig. 3 for our squares (and rectangles). It
increases roughly as $L^z, \; z \simeq 1.8$. However, a slight curvature
suggests that asymptotically the exponent $z$ may be 2. Indeed, a plot
of $\tau/L^2$ versus $1/L^{0.3}$ (Fig. 4) seems to
approach a finite limit for $L \rightarrow \infty$. This means
that when the tallest pillar hits the top, a finite fraction of
the whole $L \times L$ lattice is occupied by the bricks of the
pillars (or by the invading fluid). About the same intercept 0.07 was also
found from the high and the flat rectangles of Fig. 1, for $\tau/(L_zL)$
versus $1/L^{0.3}$.

As a function of time $t$, the average height $H$ trivially always increases 
linearly in time, while the width $W \propto t^\beta$ of the height distribution
has an exponent $\beta$ increasing towards unity for $t$ increasing towards 
$\tau$ (not shown).  

Now an interaction between neighboring sites $i$ is introduced. If
$i$ is updated, then also $i+1$ and $i-1$ are updated if
$|h_i-h_{i+1}|$ or $|h_i-h_{i-1}|$, respectively, are $\ge m$.
Here $m$ is a fixed number between 0 and $L_z$. This interaction
corresponds to some sort of surface tension, which tries to avoid
too large gradients in the height profile $h_i$. It also makes our
model one-dimensional, since now the neighborhood introduces a
geometry. The limit $m=0$ corresponds to the Bak-Sneppen model
(always updating of neighbors) and the limit $m=L_z$ to the
simplified Yee version (no updating of neighbors).

Figure 5 shows how the time $\tau$, the average height $H =
<h_i>_i$ and the surface roughness $<(h_i-H)^2>^{1/2}$ depend on
this new parameter $m$; they go neither to infinity nor to zero,
but the height $H$ has a pronounced minimum at small $m$. Because
of this new length $m$, the above simple scaling in terms of
$h/L_z$ no longer works, even if as in Fig. 6 we take $m$ to be
that value (5 to 20) for which $H$ has a minimum. Thus the
non-interacting version obeys simple scaling while the interacting
version depends on the geometry (here: one-dimensional only) and
disobeys simple scaling.

As stated, many features of the Yee model were calculated
analytically by Newman \cite{SN}. Specifically, at time $t$ the
smallest random number on the ``perimeter" was shown to grow as
\begin{equation}
x(t)=\frac{t}{t+L} \quad ,
\end{equation}
approaching unity at long times. Indeed, if a ``pillar" starts
growing at a time $t \gg L$ then it will continue to grow for $n$
consecutive steps, during which the new random numbers encountered
by this ``pillar" are smaller than $x(t)$. The probability of such
an $n$-step growth was found to be exponential: 
\begin{equation}
p_n=x^{n-1}(1-x) \quad ,
\end{equation}
yielding an average step of length
\begin{equation}
\langle n \rangle =\frac{1}{1-x}=\frac{t+L}{L} \quad.
\end{equation}
Indeed, our simulations confirm ``bursts" whose length grows
linearly with time. Also, the same formalism yields
\begin{equation}
\langle (n-\langle n \rangle)^2 \rangle=\frac{t(t+L)}{L^2} \quad {\rm or} \quad
\langle (n-\langle n \rangle)^2 \rangle/\langle n \rangle^2 = \frac{t}{t+L}
\end{equation}
which also agrees with our simulations for large $L$ and $t$, see
Fig. 7. This unifractal distribution, in which $n$ scales (for $1 \ll L
\ll t$) as $n \sim t/L$, is clearly different from that
expected in other growth models mentioned in our introduction.

Note also that a given growth stops when the next random number is
larger than $x$, which happens with probability $1-x=L/(L+t)$.
Assuming that the random numbers are distributed equally between
zero and one, every site will be encountered at least once when
$1-x$ becomes smaller than $1/L$, i. e. at times of order $L^2$.
It is not clear yet if this result relates to our numerical values
for $\tau$, which were asymptotically consistent with being $\propto 
L \times L_z \propto L^2$.

Ref. \cite{SN} also derived the probability to find a growth
``run" of length $n$ at {\it any} time,
\begin{equation}
P_n =\sum_{t=0}^\infty x(t)^{n-1}(1-x(t))^2 \propto \frac{1}{n} \quad.
\end{equation}
The final height $h_i$ of a ``pillar" is a sum over such ``run"
lengths, which grow longer and longer with time. Our simulations
show that for $L_z=L$, the typical number of such runs is of order
3--5. Thus, the distribution $N(h)$ discussed above should in
principle be a convolution of distributions like $N_n$. We
evaluated several such convolutions, with a variable number of
``runs", and they all seem to converge asymptotically (for large
$h$) to $N(h) \sim 1/h$, which differs from the exponential form
found numerically in Fig. 2. At the moment we have no explanation
for this discrepancy.

 We thank P.M.C. de Oliveira for drawing our
attention on the Sneppen model, the German-Israeli Foundation for
supporting our collaboration and the supercomputer center in
J\"ulich for time on their Cray-T3E.

\begin{figure}[hbt]
\begin{center}
\includegraphics[angle=-90,scale=0.50]{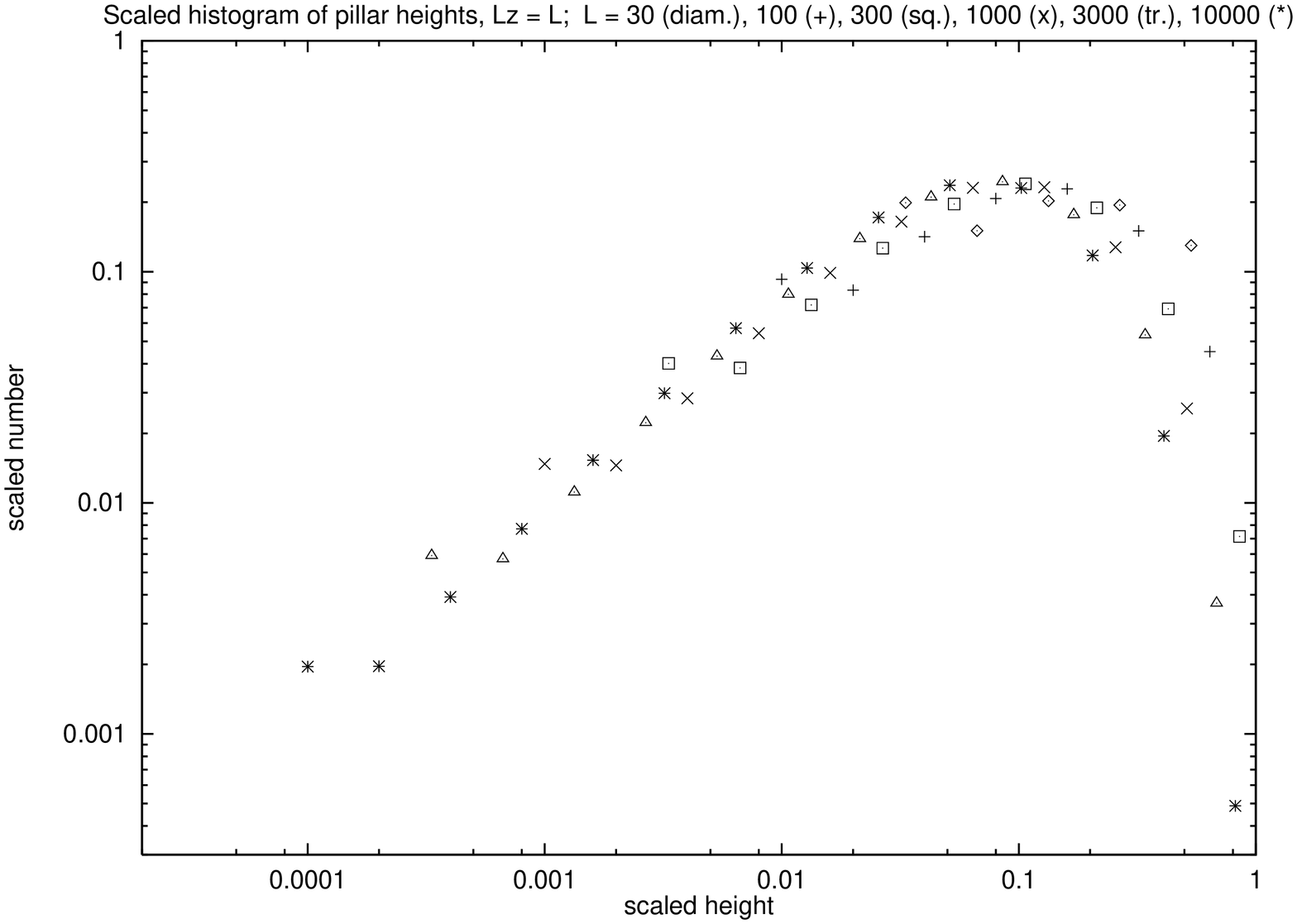}
\includegraphics[angle=-90,scale=0.25]{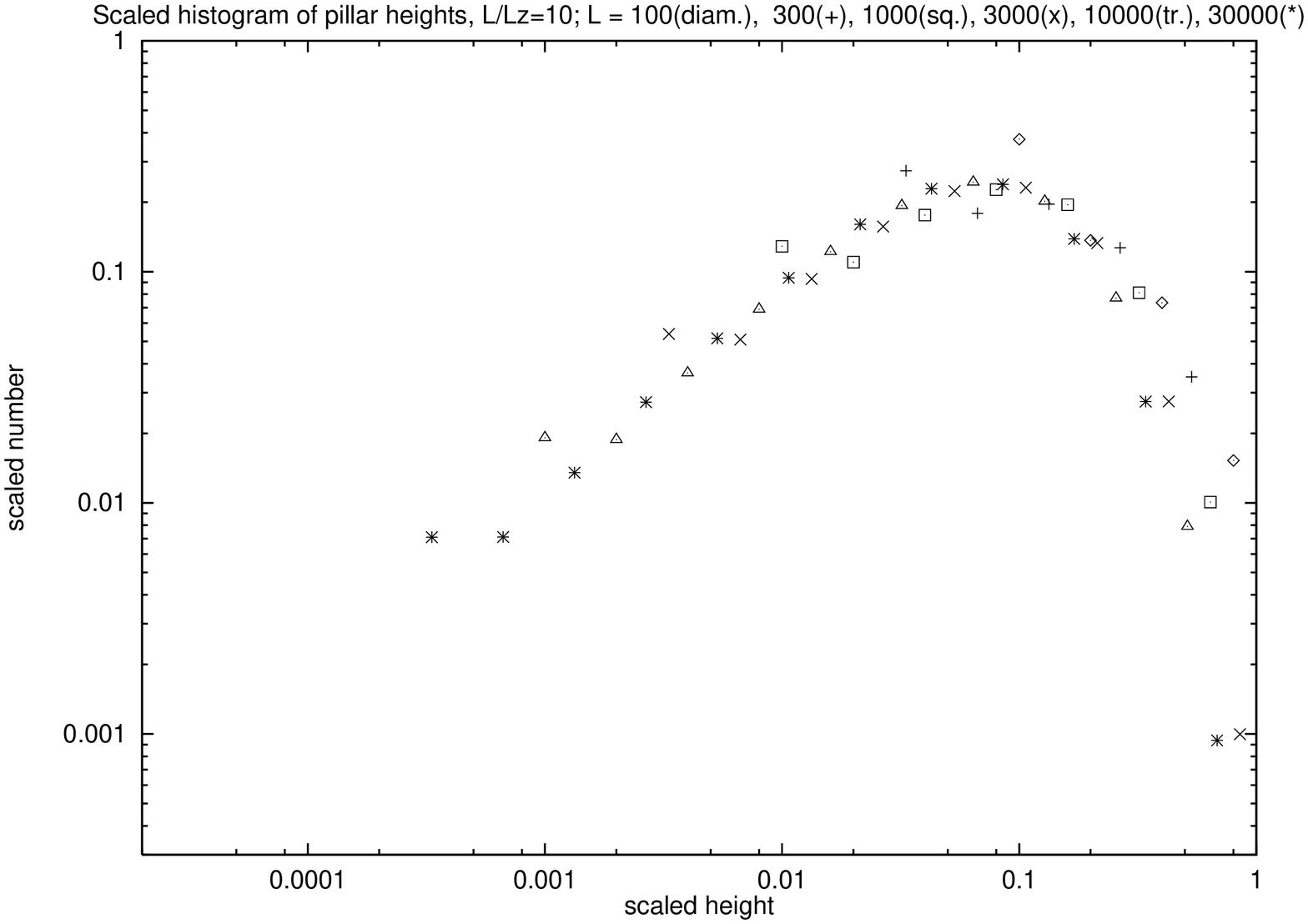}
\includegraphics[angle=-90,scale=0.25]{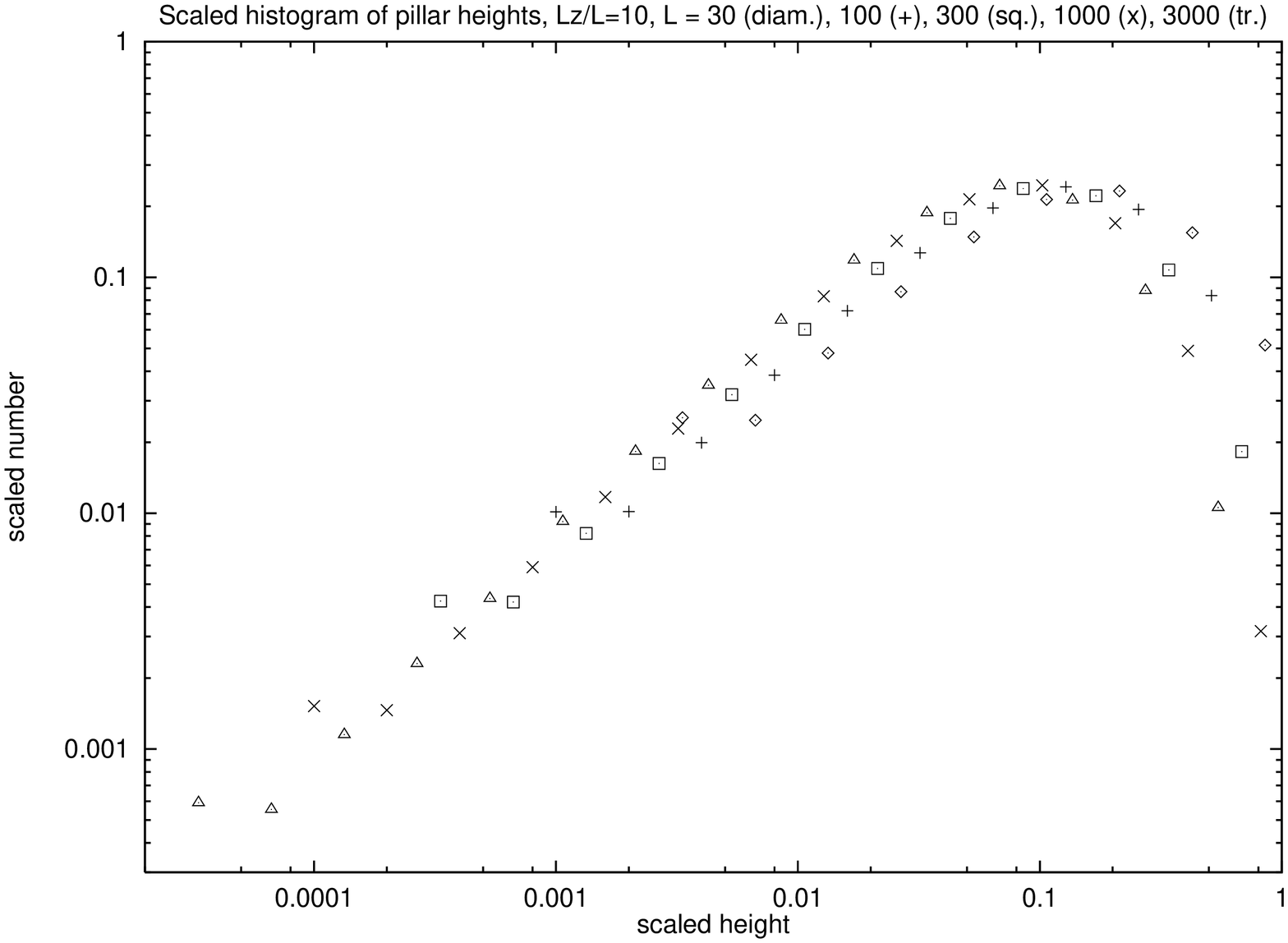}
\end{center}
\caption{
Scaling of the pillar height histograms: Curves for
different $L$ collapse, except for very small and very large
heights. The shape of the $L \times L_z$ lattice is a square (part
a), a flat rectangle (part b), and a high rectangle (part c). Up
to 640,000 samples were averaged over.
}
\end{figure}

\begin{figure}[hbt]
\begin{center}
\includegraphics[angle=-90,scale=0.5]{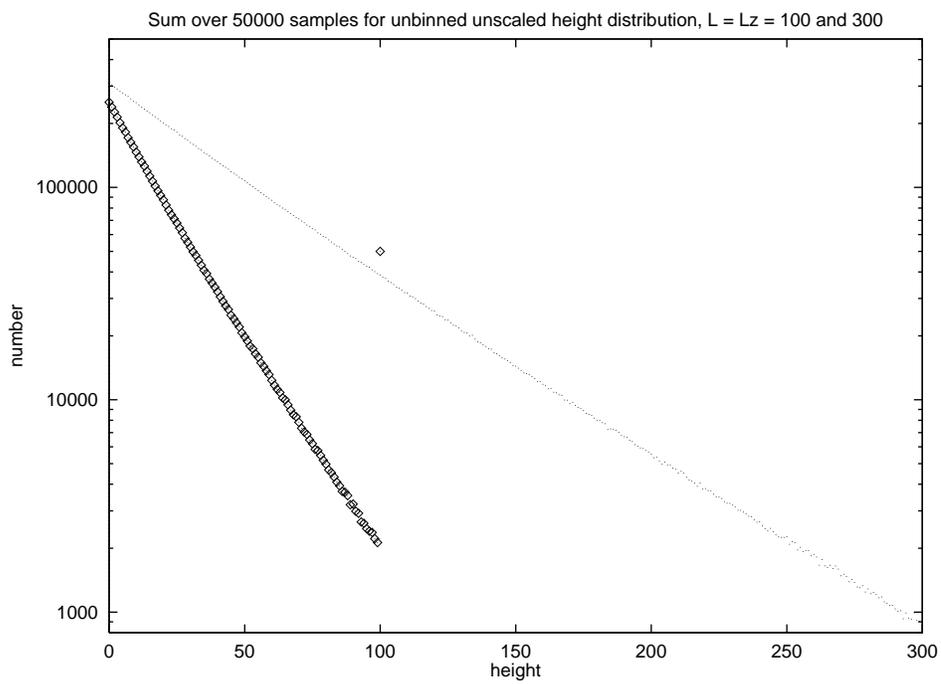}
\end{center}
\caption{
Examples of unbinned and unscaled distributions of heights, showing
a roughly exponential decay, for $L = L_z = 100$ and 300. 
}
\end{figure}

\begin{figure}[hbt]
\begin{center}
\includegraphics[angle=-90,scale=0.5]{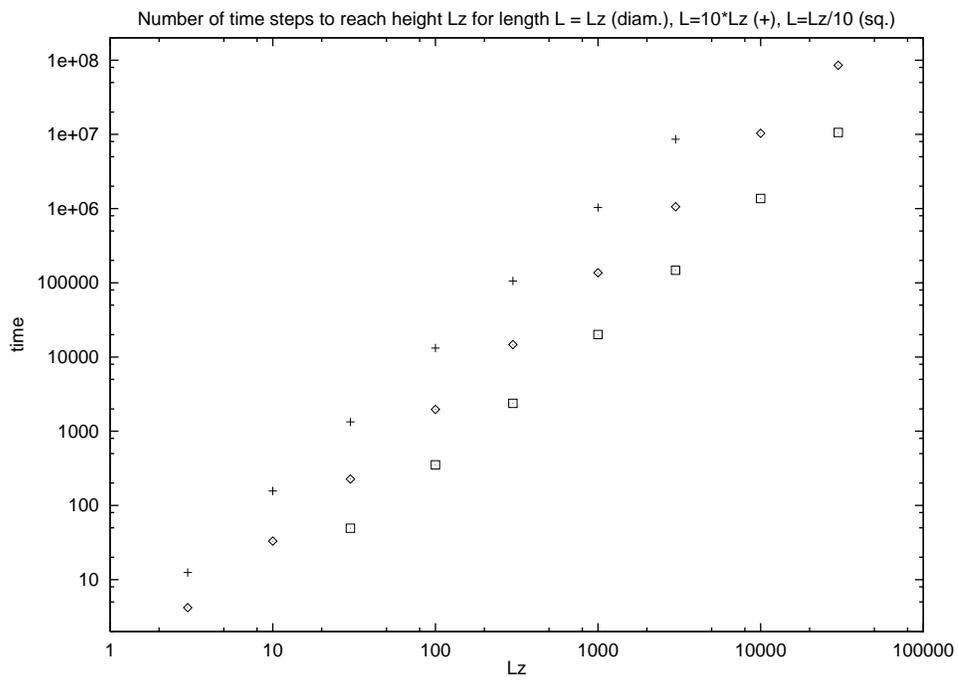}
\end{center}
\caption{
Time to reach the top, versus $L$, for squares, flat, and
high rectangles.
}
\end{figure}

\begin{figure}[hbt]
\begin{center}
\includegraphics[angle=-90,scale=0.5]{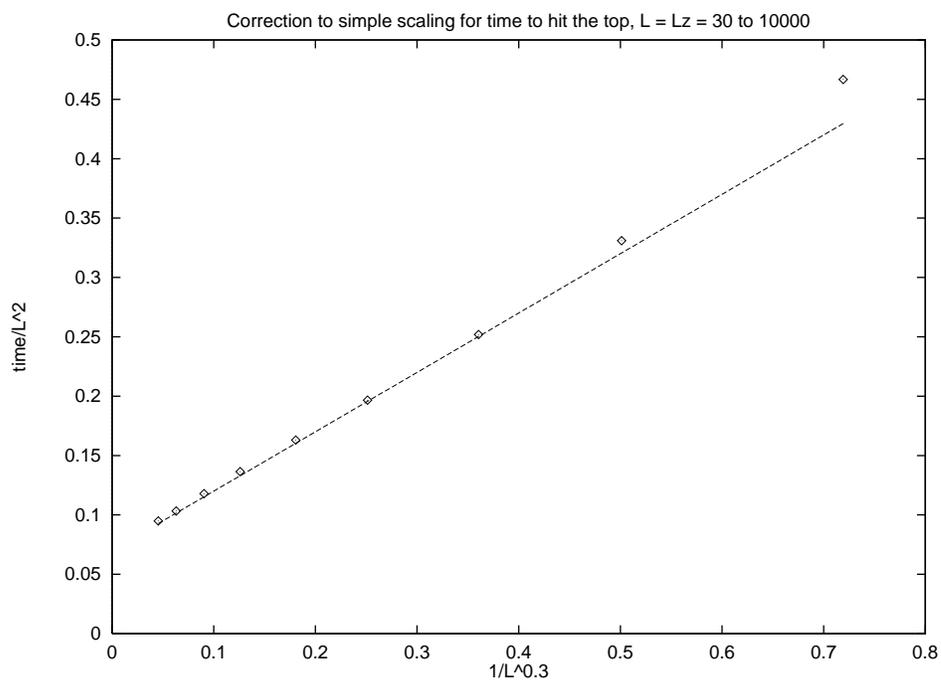}
\end{center}
\caption{
The times of Fig. 3 are shown to follow $\tau/L^2 = 0.07 + 0.5/L^{0.3}
+ \dots$ for large systems.
}
\end{figure}

\begin{figure}[hbt]
\begin{center}
\includegraphics[angle=-90,scale=0.4]{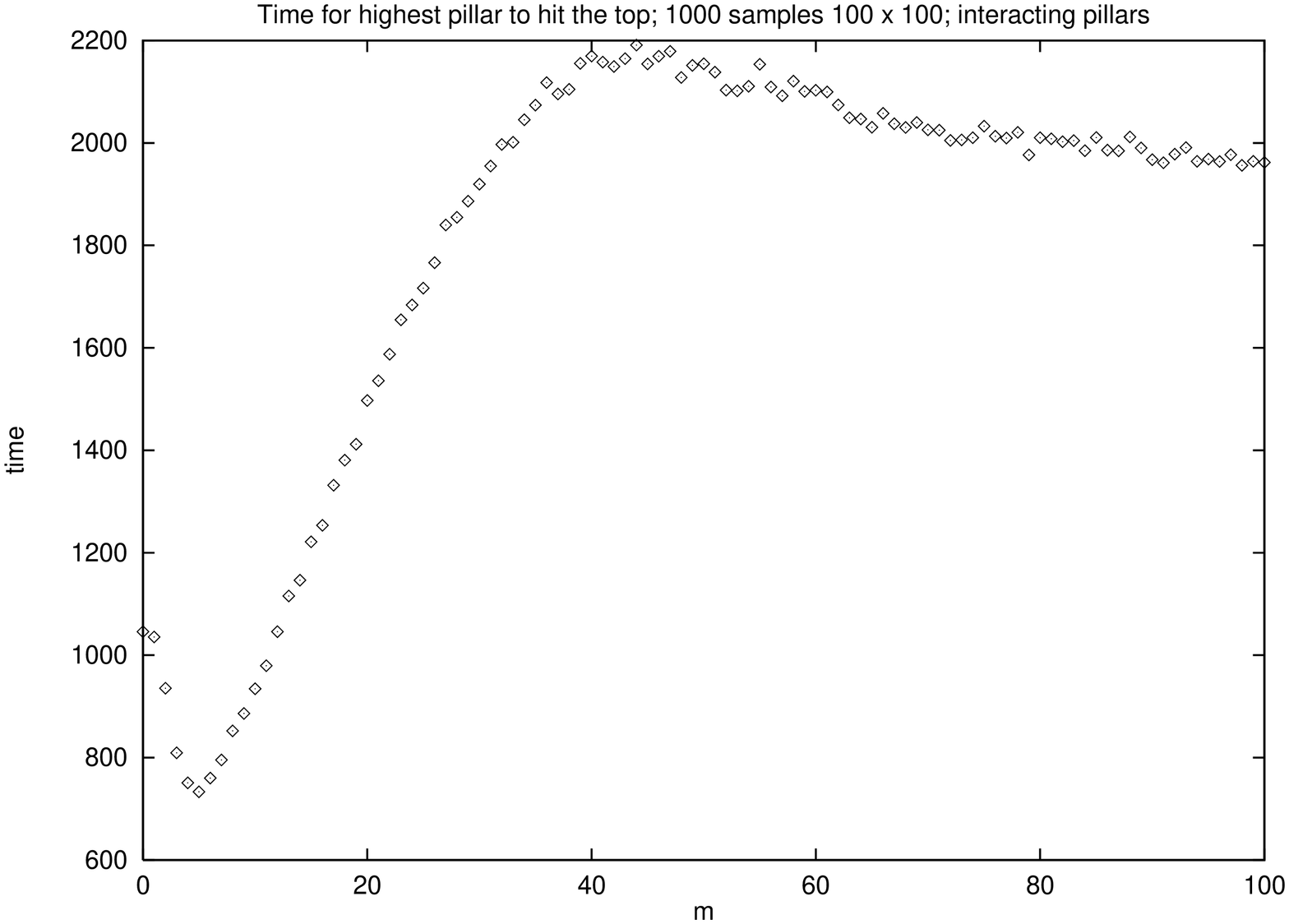}
\includegraphics[angle=-90,scale=0.4]{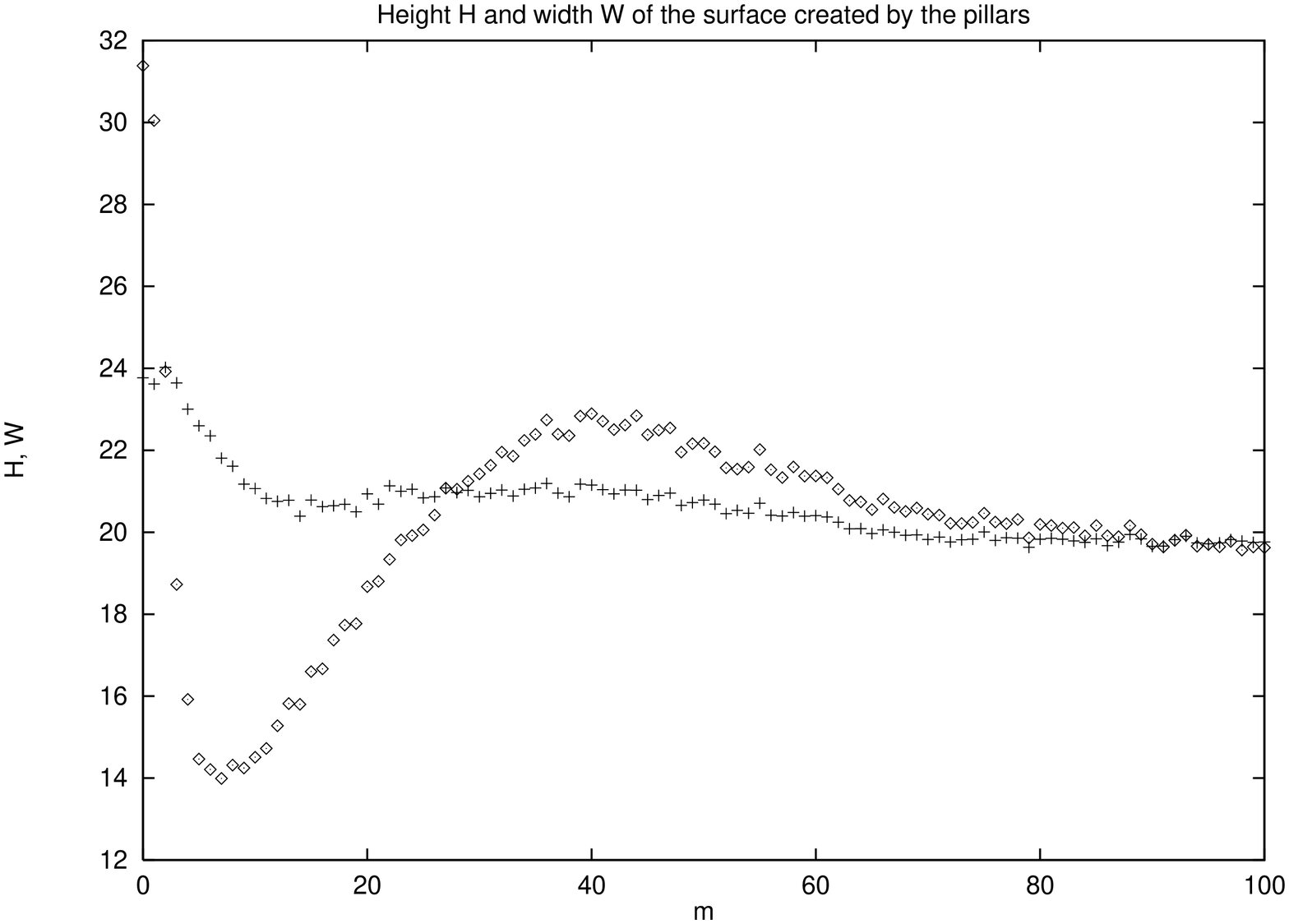}
\end{center}
\caption{
Influence of interaction parameter $m$ at $L=L_z=100$,
averaged over 1000 samples, with $m=L$ corresponding to the
simplified Yee model and $m=0$ to the one-dimensional Bak-Sneppen
model. Part a shows the time to reach the top, part b the height
(diamonds) and the width (+) of the surface defined by the pillar
tops.
}
\end{figure}

\begin{figure}[hbt]
\begin{center}
\includegraphics[angle=-90,scale=0.5]{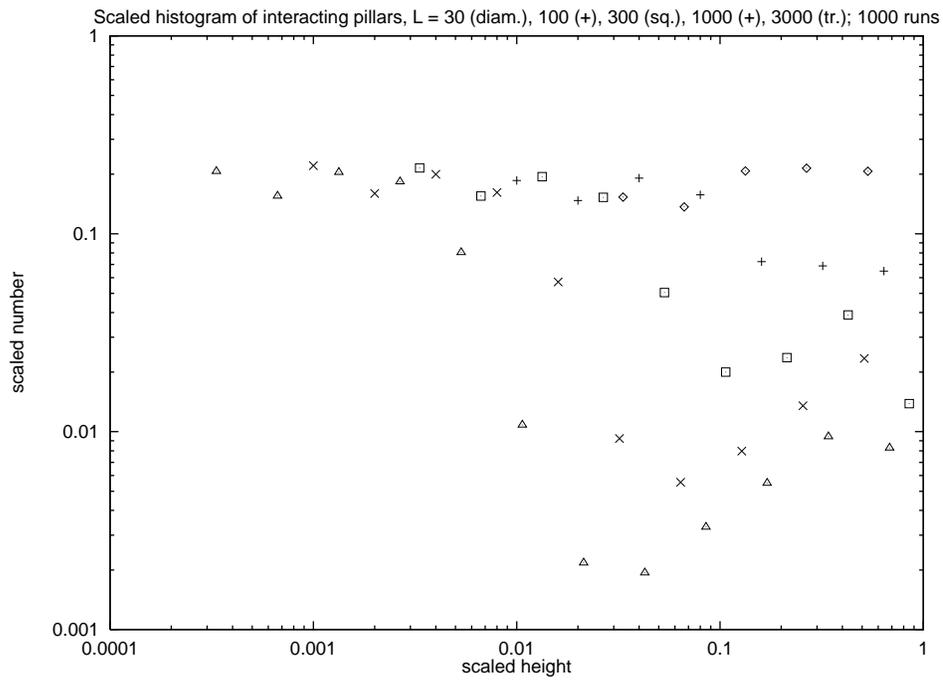}
\end{center}
\caption{
Failed scaling of pillar height histograms for thousand $L
\times L$ samples. The interaction parameter $m$ was taken to give a minimum
of $H(m)$ (diamonds in Fig. 5b) and varies from 5 to 20.
In contrast to the analogous Fig. 1, the
different curves do not collapse to one curve. 
}
\end{figure}

\begin{figure}[hbt]
\begin{center}
\includegraphics[angle=-90,scale=0.5]{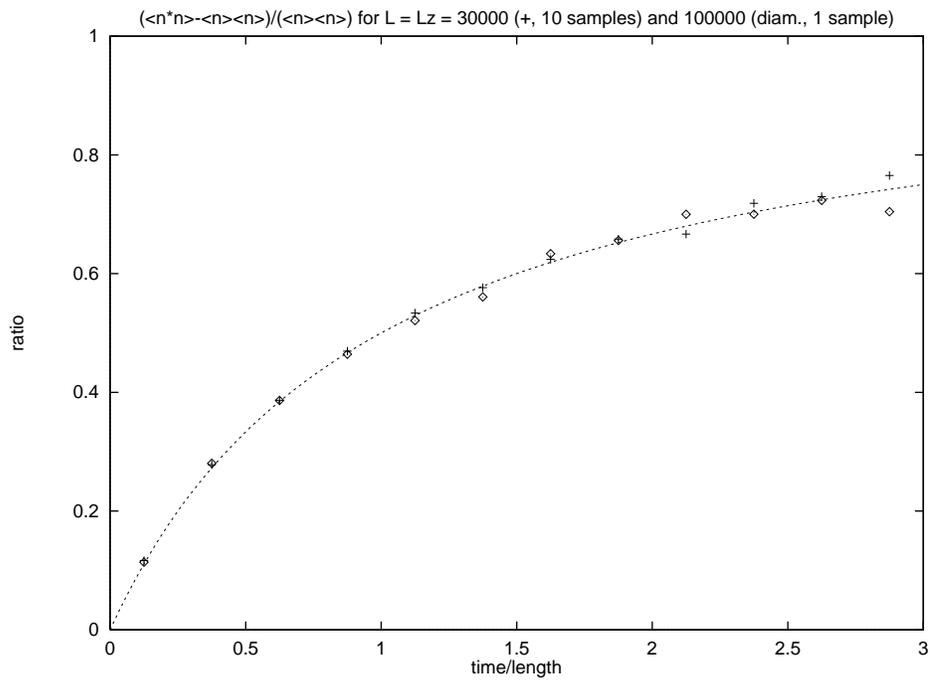}
\end{center}
\caption{
Comparison of $<(n-<n>)^2>/<n>^2$ with theoretical time
dependence $t/(t+L)$, where $n$ is the length of a stretch
(``run'' \cite{SN}) of uninterrupted updatings of the same site.
}
\end{figure}
\end{document}